%% file: main.tex
\documentclass[letterpaper, 10 pt, conference]{ieeeconf}  % Comment this line out
\IEEEoverridecommandlockouts                              % This command is only
\input{definition}

\title{\LARGE \bf
Coordinated Path Following of UAVs over Time-Varying Digraphs Connected in an Integral Sense
}

\author{Hyungsoo Kang, Isaac Kaminer, Venanzio Cichella, and Naira Hovakimyan% <-this % stops a space
\thanks{This work is supported by AFOSR, NASA, ONR and NPS CRUSER.}% <-this % stops a space
\thanks{Hyungsoo Kang and Naira Hovakimyan are with the Department of Mechanical Science and Engineering, University of Illinois at Urbana-Champaign, 
        Urbana, IL 61801, USA.  
        {\tt\small \{hk15, nhovakim\} @illinois.edu}}
\thanks{Isaac Kaminer is with the Department of
Mechanical and Aerospace Engineering, Naval Postgraduate School, Monterey, CA 93943, USA.
        {\tt\small kaminer@nps.edu}}
\thanks{Venanzio Cichella is with the Department of Mechanical
Engineering, University of Iowa, 
        Iowa City, IA 52242, USA.
        {\tt\small venanzio-cichella@uiowa.edu}}  
}

\begin{document}

\maketitle
\thispagestyle{empty}
\pagestyle{empty}

%%%%%%%%%%%%%%%%%%%%%%%%%%%%%%%%%%%%%%%%%%%%%%%%%%%%%%%%%%%%%%%%%%%%%%%%%%%%%%%%
\begin{abstract}
This paper presents a new connectivity condition on the information flow between UAVs to achieve coordinated path following. The information flow is directional, so that the underlying communication network topology is represented by a time-varying digraph. We assume that this digraph is connected in an integral sense. This is a much more general assumption than the one currently used in the literature. Under this assumption, it is shown that a distributed coordination controller ensures exponential convergence of the coordination error vector to a neighborhood of zero. The efficacy of the algorithm is confirmed with simulation results.

%This paper presents a new connectivity condition on the information flow between quadrotors to achieve coordinated path following. The information flow is no longer required to be bidirectional, so its topology is a time-varying digraph. Connectedness of it in an integral sense is a weaker condition than the previous ones. Under this condition, it is shown that a decentralized coordination controller ensures convergence of the coordination error to a neighborhood of zero. 

%This paper presents a time-coordination algorithm designed on the basis of the condition that the directed inter-UAV information flow is connected in an integral sense. It is a sufficiently weak connectivity condition, which can be satisfied when the topology is not connected for some length of time or even when not connected at all times. Thus, the algorithm can attain the coordination objectives with switching topologies and also is robust to loss of communication links.% due to switching topology or loss of communication.
\end{abstract}

%%%%%%%%%%%%%%%%%%%%%%%%%%%%%%%%%%%%%%%%%%%%%%%%%%%%%%%%%%%%%%%%%%%%%%%%%%%%%%%%
\section{INTRODUCTION}
%Coordination of multiple quadrotors has been an active area of research over the past two decades due to its wide range of applications. Well known examples are

Over the past two decades, we have witnessed significant progress in theories and technologies on multi-agent coordination. %Multi-agent systems have been widely used in our life. 
Presently, multi-agent systems are widely used for a wide variety of missions. Representative examples include multi-UAV collaborative SLAM \cite{Schmuck2017}, \cite{Dubé2017}, surveillance missions \cite{Peters2017}, \cite{Botts2016}, cooperative payload transportation \cite{Lee2017}, and wild fire tracking \cite{Pham2020}.

Among diverse research results in the field, coordinated path-following algorithms have been developed to provide solutions to challenging problems such as simultaneous suppression of multiple targets located at different positions, sequential auto-landing, and flying in formation to name a few. The key feature of all these missions is  that all the UAVs must arrive at their destinations at the same time.  To achieve this objective the coordinated path-following framework includes  trajectory generation, path-following and time-coordination  control. At the trajectory generation step, the trajectory generation algorithm \cite{Choe2016}, \cite{Cichella2021} computes a set of desired trajectories that minimize a cost function subject to constraints such as simultaneous arrival, boundary conditions, safety distances between vehicles and other obstacles, and dynamic constraints of the UAVs.  Given a desired trajectory obtained in the previous step, the path-following controller \cite{Cichella2013} has each UAV track the desired trajectory. 
%The important limitation of the path-following controller is that it has a finite region of attraction. When a UAV is pushed ahead or behind by disturbances, the desired position needs to speed up or slow down its progression along the trajectory to put the UAV inside the region of attraction, which definitely help us not lose the quadrotor. However, it comes at the cost of discoordination between the quadrotors. 
However, in the presence of disturbances the path following controllers cannot guarantee simultaneous arrival by all the UAVs at the end of their respective trajectory. 
This is despite the fact that the trajectories were designed for all the UAVs to arrive simultaneously at their final destinations.  

%once the fleet is discoordinated by disturbances, there is no way to recover the coordination only with the path-following controller. Eventually, they will arrive at the final destinations at different times.

To resolve this issue researchers came up with distributed time-coordination algorithms that exchange information on the progress of each UAV along its trajectory over an underlying communication network.
%In this approach, each UAV exchanges information with its neighbors on the progress it made along its trajectory  and adjusts its progression speed referring to neighbors' states. 
This implies that the type of the inter-vehicle communication has a major impact on the simultaneous arrival performance by all the UAVs. As the time-coordination objective is not only the state consensus $\gamma_i(t)=\gamma_j(t)$, $\forall i,j$ but also the state rate consensus $\dot{\gamma}_i(t)=\dot{\gamma}_d(t)$, $\forall i$, where $\dot{\gamma}_d(t)$ is a given reference function, the numerous existing consensus algorithms designed only for the state consensus over bidirectional/directed graphs cannot be applied straightforwardly. The papers \cite{Ghabcheloo2007,Xargay2013,Cichella2015} presented connectivity conditions and distributed controllers to solve the time-coordination problem. To be specific, the authors in~\cite{Ghabcheloo2007} worked out an algorithm with the assumption that the topology of the information flow is connected via a bidirectional graph. This assumption was weakened in  \cite{Xargay2013}, \cite{Cichella2015} as follows: the network topology was assumed to be represented by a time-varying bidirectional graph such that the integral of the  graph from $t$ to $t+T$ is connected $\forall t\geq 0$ with $T>0$, i.e., connected in an integral sense. With this condition, it was shown in \cite{Mehdi2019}, \cite{Tabasso2021} that collision avoidance can be achieved as well as the time-coordination. The authors in  \cite{Puig-Navarro2015} proposed a time-coordination control law with absolute temporal constraints. Importantly, all these algorithms were designed on the basis of a restrictive assumption: the information flow across the network must be bidirectional. A natural question to ask is whether a weaker network connectivity assumption exists, for example, when the information flow takes place over a time-varying digraph? Is it possible to guarantee simultaneous arrival in this case? These have been long-standing challenging questions because the existing time-coordination algorithms \cite{Xargay2013,Cichella2015,Mehdi2019,Tabasso2021,Puig-Navarro2015} were designed using the symmetry of the Laplacian of a bidirectional network. It implies that they could not be easily extended to the digraph case where the Laplacian is not symmetric. 
%Furthermore, the time-coordination objectives include not only the state consensus $\gamma_i(t)=\gamma_j(t)$ but also the state rate consensus $\dot{\gamma}_i(t)=\dot{\gamma}_d(t)$, where $\dot{\gamma}_d(t)$ is a reference function. Due to the ambitious objectives, numerous existing state-only consensus algorithms over time-varying digraphs cannot be applied to our problem. This motivated our research. 
To address this issue, the authors' previous work \cite{Kang2021} derived a switching rule for the network topology. It showed that the time-coordination objectives can be achieved when the topology takes a certain one of jointly connected digraphs $\mathcal{D}_1$, $\dots$, $\mathcal{D}_n$ at a certain time in accordance with the derived switching rule. However, this is an artificial and confining constraint on the network topology. Therefore, it is more desirable to derive a reasonably weak connectivity condition for a time-varying digraph $\mathcal{D}(t)$.

The contributions of this paper are summarized as follows. The inter-vehicle communication is no longer required to be bidirectional, i.e. the topology of the underlying network is assumed to be represented by a time-varying digraph. Moreover, we assume that the time-varying digraph is connected in an integral sense.
%which is more general than the assumption used in \cite{7065327} that the time-varying bidirectional graph is connected in an integral sense. 
We then employ Lyapunov analysis to show that a proposed decentralized time-coordination control law guarantees exponential convergence of the time-coordination error vector to a neighborhood of zero. This implies that all the UAVs arrived at their final destinations at the same time. The proposed time-coordination algorithm works for any agent endowed with a path-following controller. In this paper, we illustrate performance of the proposed algorithm using an example of the coordinated path following of quadrotors borrowed from \cite{Cichella2013}.
%because quadrotors are one of the most widely used unmanned vehicles in our life. 

The rest of this article is organized as follows. Section~\ref{prelim} provides the basics of graph theory. Section~\ref{III} formulates the coordinated path-following problem addressed in this paper. Section~\ref{IV} proposes a time-coordination algorithm that solves the problem. Section~\ref{V} includes simulation results. Finally, Section~\ref{VI} presents conclusions.
%%%%%%%%%%%%%%%%%%%%%%%%%%%%%%%%%%%%%%%%%%%%%%%%%%%%%%%%%%%%%%%%%%%%%%%%%%%%%%%%
\section{PRELIMINARIES} \label{prelim}
\subsection{Graph Theory}
A digraph is denoted by $\mathcal{D}=(\mathcal{V},\mathcal{E},\mathcal{A})$, where $\mathcal{V}$ is the set of nodes, $\mathcal{E}$ is the set of edges, and $\mathcal{A}$ is the Adjacency matrix. A directed edge $(i,j)$ implies that information can be transmitted from node $j$ to node $i$. 
If $(i,j)\in \mathcal{E}$, one has $\mathcal{A}_{ij}=1$. Otherwise, $\mathcal{A}_{ij}=0$. The digraph $\mathcal{D}$ is represented by the Laplacian $L\trieq \Delta-\mathcal{A}$, where $\Delta$ is a diagonal matrix with $\Delta_{ii}\trieq\sum_{j=1,j\neq i}^{n}\mathcal{A}_{ij}$. The neighborhood of node $i$ is $\mathcal{N}_i\trieq\{j\in\mathcal{V}: (i,j)\in \mathcal{E}\}$. A directed path from a node $i_s$ to $i_0$ is a sequence of edges $(i_0,i_1)$, $(i_1,i_2)$, $\dots$, $(i_{s-1},i_s)$. If there exists a node such that every other node is reachable via a directed path, the digraph is said to contain a directed spanning tree. 

For a time-varying digraph $\mathcal{D}(t)$, let us consider the digraph represented by the integrated Laplacian $\int_{t}^{t+T}L(\tau)d\tau$. An edge $(i,j)$ in it is said to be a $\delta$-edge if $\int_{t}^{t+T}-L_{ij}(\tau)d\tau\geq\delta$. A path in it is said to be a $\delta$-path if every edge on the path is a $\delta$-edge.
%\subsection{Notation}

%%%%%%%%%%%%%%%%%%%%%%%%%%%%%%%%%%%%%%%%%%%%%%%%%%%%%%%%%%%%%%%%%%%%%%%%%%%%%%%%

\section{TIME-COORDINATED PATH-FOLLOWING FRAMEWORK} \label{III}
\subsection{Path Following of a Single UAV}
The trajectory generation algorithm in \cite{Choe2016}, \cite{Cichella2021} designs a set of $n$ collision-free desired trajectories
\begin{align} \label{trajectory}
    p_{d,i}(t_d): [0,t_f]\rightarrow \mathbb{R}^3, \ \ i\in\{1,\dots,n\},
\end{align}
where $t_f$ is the mission duration. Let us denote the desired position of the $i$th UAV as $p_{d,i}(\gamma_i(t))$, where $\gamma_i(t)$ is an adjustable nondecreasing variable
\begin{align*}
    \gamma_i(t): [0,\infty)\rightarrow[0,t_f], \ \ \ i\in\{1,\dots,n\}, 
\end{align*}
called virtual time or coordination state. It is evident that adjustment of $\gamma_i(t)$ regulates the progression speed of the desired position of the UAV along the trajectory. The algorithm that adjusts $\gamma_i(t)$'s to coordinate $n$ UAVs will be presented in section~\ref{IV}.

Notice that the UAV needs a path-following control law in order to steer its actual position $p_i(t)$ to the desired position $p_{d,i}(\gamma_i(t))$. The geometric path-following controller introduced in \cite{Cichella2013} ensures that the norm of the path following error
\begin{align*}
    e_{PF,i}(t)\trieq p_i(t)-p_{d,i}(\gamma_i(t)), \ \ \ i\in\{1,\dots,n\} 
\end{align*}
converges exponentially to zero with the ideal performance of the autopilot and to a neighborhood of zero with non-ideal one.

\subsection{Time Coordination of Multiple UAVs} 
As seen in the previous subsection, the coordination state $\gamma_i(t)$ characterizes the progression of each UAV along a desired trajectory. We say that the UAVs are synchronized at time $t$, if 
\begin{align} \label{obj1}
    \gamma_i(t)=\gamma_j(t), \ \ \ \forall i,j\in\{1,\dots,n\}.
\end{align}
Also, the UAVs are considered to progress at the same desired pace, if
\begin{align} \label{obj2}
    \dot{\gamma}_i(t)=\dot{\gamma}_d(t), \ \ \ \forall i\in\{1,\dots,n\},
\end{align}
where $\dot{\gamma}_d(t)$ is the desired mission pace. 

In order to attain the above coordination objectives, each UAV needs to transmit its coordination state as well as adjusting with respect to the coordination states transmitted from the neighboring UAVs. Clearly, the information flow between the UAVs plays an important role in the process of achieving the goal of simultaneous arrival.
%The information flow over can be rigorously described by using graph theory. 
In this paper we make the following assumptions on the digraph that models the information flow between the UAVs:
\begin{assumption} \label{assum1}
The information flow between two UAVs is directional with no time delays.
\end{assumption}

\begin{assumption} \label{assum2}
The $i$th UAV can receive coordination information $\gamma_j(t)$ only %from its neighboring UAVs in $\mathcal{N}_i(t)$.
from UAVs in its neighborhood set $\mathcal{N}_i(t)$, where $j\in\mathcal{N}_i(t)$.
\end{assumption}

\begin{assumption} \label{assum3}
For all $t\geq0$, there exists $T>0$ such that the digraph represented by $\int_{t}^{t+T}L(\tau)d\tau$ contains a $\delta$-spanning tree. That is, a root node in it can reach every other node via a $\delta$-path. 
\end{assumption}

\begin{remark} 
The parameters $T>0$ and $\delta \in (0,T]$ represent the Quality of Service (QoS) of the underlying network. A smaller value of $T$ and a value of $\delta$ closer to $T$ indicates stronger connectivity.
\end{remark}

\begin{remark}
We note that Assumption~\ref{assum3} requires the information flow to be connected in an integral sense, not pointwise in time. It is a reasonably weak connectivity condition since it can be satisfied when the network is not connected during a certain portion of the mission or even when it fails to be connected at all times.  
\end{remark}

\begin{remark}
    In \cite{Cichella2015}, the connectivity conditions for achieving the coordination objectives are 1) inter-vehicle communication is bidirectional, 2) $\int_{t}^{t+T}L(\tau)d\tau$ is a connected bidirectional graph. If these two conditions are satisfied, then  Assumptions~\ref{assum1} and \ref{assum3}, respectively, are also satisfied. However, the inverse is not true. In this regard, the results presented in this paper are a generalization of the ones in \cite{Cichella2015}. %with generalized and weakened connectivity assumptions.
\end{remark}

\noindent \textit{TCPF Problem (Time-Coordinated Path-Following Problem):} Given a set of desired trajectories \eqref{trajectory} and a path-following control law that ensures $\|e_{PF,i}(t)\|$ converges to a neighborhood of zero exponentially, design a distributed time-coordination controller such that $|\gamma_i(t)-\gamma_j(t)|$ and $|\dot{\gamma}_i(t)-\dot{\gamma}_d(t)|$ converge to a neighborhood of zero under Assumptions~\ref{assum1}, \ref{assum2}, and \ref{assum3}.
%%%%%%%%%%%%%%%%%%%%%%%%%%%%%%%%%%%%%%%%%%%%%%%%%%%%%%%%%%%%%%%%%%%%%%%%%%%%%%%%
\section{DISTRIBUTED TIME-COORDINATION CONTROLLER : MAIN RESULT} \label{IV}
We propose the following distributed time-coordination control law
%as in \cite{7065327} 
to solve the TCPF Problem under Assumptions \ref{assum1} and \ref{assum2}
\begin{align} \label{dyn1}
    \ddot{\gamma}_i(t)&=-b(\dot{\gamma}_i(t)-\dot{\gamma}_d(t)) \nonumber \\
    &\mathrel{\phantom{=}}-a\sum_{j\in\mathcal{N}_i(t)}(\gamma_i(t)-\gamma_j(t))+\bar{\alpha}_i(e_{PF,i}(t)), \\
    \gamma_i(0)&=\gamma_{i0}, \ \ \dot{\gamma}_i(0)=\dot{\gamma}_{i0}, \nonumber
\end{align}
where $a$ and $b$ are positive coordination control gains and $\bar{\alpha}_i(e_{PF,i}(t))$ is defined as 
\begin{align} \label{alpha}
    \bar{\alpha}_i(e_{PF,i}(t))=\frac{\dot{p}_{d,i}(\gamma_i(t))^\top e_{PF,i}(t)}{\|\dot{p}_{d,i}(\gamma_i(t))\|+\epsilon},
\end{align}
with $\epsilon$ being a positive design parameter. 
\begin{remark}
    The time-coordination control law \eqref{dyn1} is motivated by the work presented in \cite{Cichella2015}. 
\end{remark}
\begin{remark}
    The term $\bar{\alpha}_i(e_{PF,i}(t))$ is designed to take into consideration that the geometric path-following controller \cite{Cichella2013} has a finite region of attraction. If a UAV precedes (falls behind) its desired position, the numerator becomes positive (negative), thereby accelerating (decelerating) progression of the desired position. In other words, it is geared towards decreasing $\|e_{PF,i}(t)\|$, which helps the UAV to quickly approach the desired position staying inside the region of attraction.  
\end{remark}
For notational simplicity, we introduce the coordination error state $\xi_{TC}(t)=[\xi_1(t)^\top \ \xi_2(t)^\top]^\top$ with
\begin{equation}
\begin{aligned} \label{error}
    \xi_1(t)&=Q\gamma(t) \ \in \mathbb{R}^{n-1}, \\
    \xi_2(t)&=\dot{\gamma}(t)-\dot{\gamma}_d(t)1_n \ \in \mathbb{R}^{n}, \nonumber
\end{aligned}
\end{equation}
where $\gamma(t)=[\gamma_1(t), \dots, \gamma_n(t)]^\top$ and $Q\in\mathbb{R}^{(n-1)\times n}$ is a matrix that satisfies $Q1_n=0_{n-1}$ and $QQ^\top=\mathbb{I}_{n-1}$. 
\begin{remark}
    A matrix $Q_k\in\mathbb{R}^{(k-1)\times k}$, $k\geq2$ satisfying $Q_k1_k=0_{k-1}$ and $Q_k\left(Q_k\right)^\top=\mathbb{I}_{k-1}$ can be constructed recursively:
    \begin{align*}
        Q_k=
        \begin{bmatrix}
        \sqrt{\frac{k-1}{k}} & -\frac{1}{\sqrt{k(k-1)}}1_{k-1}^\top \\
        0 & Q_{k-1} \\
        \end{bmatrix}
    \end{align*}
    with initial condition $Q_2=[1/\sqrt{2} \ -1/\sqrt{2}]$. For notational simplicity, we denote $Q_n$ by $Q$, where $n$ is the number of  UAVs.
\end{remark}
%\noindent According to Lemma $7$ in \cite{phdenric2013}, it holds that $Q^\top Q=\mathbb{I}_n-\frac{1_n1_n^\top}{n}$
%\\ \\ 
It is shown in \cite[Lemma~7]{Phd_enric2013} that $Q^\top Q=\mathbb{I}_n-\frac{1_n1_n^\top}{n}$ and the nullspace of Q is spanned by $1_n$. Therefore, if $\xi_1(t)=Q\gamma(t)=0_{n-1}$, then it holds that $\gamma_i(t)=\gamma_j(t)$, $\forall i,j\in\{1,\dots,n\}$. Furthermore, $\xi_2(t)=0_n$ implies that 
$\dot{\gamma}_i(t)=\dot{\gamma}_d(t)$, $\forall i\in\{1,\dots,n\}$. Therefore, $\xi_{TC}(t)=0_{2n-1}$ is equivalent to \eqref{obj1} and \eqref{obj2}. 

Using the notation introduced in the previous paragraphs, the dynamics of $\gamma(t)$ can be rewritten as 
\begin{align} \label{collective_dyn}
    \ddot{\gamma}(t)&=-b\xi_2(t)-aL(t)\gamma(t)+\bar{\alpha}(e_{PF}(t)), \\
    \gamma(0)&=\gamma_0, \ \ \dot{\gamma}(0)=\dot{\gamma}_0, \nonumber
\end{align} 
where $\bar{\alpha}(e_{PF}$ $(t))=[\bar{\alpha}_1(e_{PF,1}(t)),\dots,\bar{\alpha}_n(e_{PF,n}(t))]^\top$.

%\begin{lemma} \label{lem1}
%Define $\bar{L}(t)\trieq QL(t)Q^\top \in \mathbb{R}^{(n-1)\times(n-1)}$. Then, the following hold at any time $t$. \\
%a) The spectrum of $\bar{L}(t)$ is the same as that of $L(t)$ without the eigenvalue $0$ whose corresponding eigenvector is $1_n$. \\
%b) If $\mathcal{D}(t)$ contains a spanning tree, $-\bar{L}(t)$ is~Hurwitz stable. Otherwise, $-\bar{L}(t)$ is marginally stable.
%\end{lemma}
%\begin{proof}
%a) Given $L(t)x=\lambda x$ $(x\neq 0)$, one has $QL(t)x=\lambda Qx$. 
%The left hand side of the latter equation is $QL(t)x=QL(t)\left(\mathbb{I}_n-\frac{1_n1_n^\top}{n}\right)x=QL(t)Q^\top Qx=\bar{L}(t)Qx$.
%Further, b) is deduced from a) and the algebraic connectivity of digraphs presented in section \ref{prelim}. 
%\end{proof} 

\begin{lemma} \label{lem1}
Consider the following dynamics
\begin{align} \label{lem1:equ1}
    \dot{x}=-\frac{a}{b}L(t)x, \ \ \ x(0)=x_0\in\mathbb{R}^n,
\end{align}
where $a$ and $b$ are the positive coordination control gains in \eqref{dyn1}. Under Assumption~\ref{assum3} on $L(t)$, the components $x_i$'s of $x$ reach consensus exponentially
\begin{equation}
\begin{aligned}
    diam\left(x\left(t\right)\right)&\trieq\max\limits_{i}\{x_i(t)\}-\min\limits_{i}\{x_i(t)\} \\
    &\leq diam\left(x(0)\right)ke^{-\lambda t}, \label{lem1:equ2}
\end{aligned}
\end{equation}
where $k\trieq\frac{1}{1-(\delta')^n}$ and $\lambda\trieq-\frac{1}{nT}\ln(1-(\delta')^n)$ with $\delta'\trieq \min\left\{1,\frac{a}{b}\delta\right\}e^{-(n-1)\frac{a}{b}T}$.
\end{lemma}
\begin{proof}
    The proof is similar to the first half of the proof of \cite[Theorem~1]{Han2015}.
\end{proof}

\begin{lemma} \label{lem2}
The quantities $\|Qx\|$ and $diam(x)$ satisfy
%represent discoordination among $x_i$'s in a vector $x$, are in the following relation:
\begin{align} \label{lem2:equ}
    \frac{1}{\sqrt{n}}\|Qx\|\leq diam(x)\leq \sqrt{2}\|Qx\|
\end{align}
\end{lemma}
\begin{proof} 
The first inequality is proven by considering
\begin{align*}
    \resizebox{\hsize}{!}{$\|Qx\|^2=x^\top Q^\top Qx=x^\top \left(\mathbb{I}_n-\frac{1_n1_n^\top}{n}\right) x=n\left\{\frac{1}{n}\sum_{i=1}^{n}x_i^2-\left(\frac{1_n^\top x}{n}\right)^2\right\}$} \\
    \resizebox{\hsize}{!}{$=\sum_{i=1}^{n}\left(x_i-\frac{1_n^\top x}{n}\right)^2\leq n\left(\max\limits_{i}\{x_i\}-\min\limits_{i}\{x_i\}\right)^2=n\left\{diam(x)\right\}^2$}.
\end{align*}
To prove the second inequality, note that
\begin{align*}
    \resizebox{\hsize}{!}{$\left\{diam(x)\right\}^2=\left(\max\limits_{i}\{x_i\}-\min\limits_{i}\{x_i\}\right)^2=\left(\max\limits_{i}\{x_i\}-\frac{1_n^\top x}{n}+\frac{1_n^\top x}{n}-\min\limits_{i}\{x_i\}\right)^2$} \\
    \resizebox{\hsize}{!}{$\leq2\left(\max\limits_{i}\{x_i\}-\frac{1_n^\top x}{n}\right)^2+2\left(\frac{1_n^\top x}{n}-\min\limits_{i}\{x_i\}\right)^2\leq2\sum_{i=1}^{n}\left(x_i-\frac{1_n^\top x}{n}\right)^2=2\|Qx\|^2$}.
\end{align*}
\end{proof} 
\begin{remark}
When the components $x_i$'s of a vector $x$ reach consensus, i.e., $x_1=\cdots=x_n$, both $\|Qx\|$ and $diam(x)$ are zero. Otherwise, they are positive numbers, which become larger as $x_i$'s diverge from consensus. Therefore, we can say that $\|Qx\|$ and $diam(x)$ measure discoordination among $x_i$'s. We note that the relation \eqref{lem2:equ} indicates that $\|Qx\|$ and $diam(x)$ are equivalent measures.
\end{remark}
The following theorem introduces the main result of this article.
\begin{theorem} \label{thm}
Consider a set of desired trajectories \eqref{trajectory} and a path-following controller that ensures $\|e_{PF,i}(t)\|$ converges to a neighborhood of zero exponentially. Let the evolution of $\gamma_i(t)$ be governed by \eqref{dyn1} under Assumptions~\ref{assum1}, \ref{assum2}, and \ref{assum3}. Then, there exist time coordination control gains $a$, $b$, and $\epsilon$ such that
\begin{equation}
\begin{aligned}
    \|\xi_{TC}(t)\|\leq\kappa_1\|\xi_{TC}&(0)\|e^{-\lambda_{TC}t} \\&+\kappa_2\sup_{t\geq0}\left(\|e_{PF}(t)\|+|\ddot{\gamma}_d(t)|\right)
\end{aligned}
\end{equation}
with the rate of convergence
\begin{align} \label{rate}
    \lambda_{TC}\leq\frac{\lambda}{6nk^2},
\end{align}
where $\lambda$ and $k$ were defined in Lemma~\ref{lem1}.
\end{theorem}
\begin{proof}
%To analyze the convergence properties of \eqref{dyn1}, motivated by \cite{7065327}, we reformulate it into a stabilization problem by introducing a variable
%For ease of Lyapunov analysis, motivated by \cite{7065327}, we reformulate \eqref{dyn1} into a stabilization problem by introducing a variable
Let
\begin{align*}
    \chi(t)=b\xi_1(t)+Q\xi_2(t).
\end{align*}
Then, the coordination error state $\xi_{TC}(t)=[\xi_1(t)^\top \ \xi_2(t)^\top]^\top$ can be redefined by $\bar{\xi}_{TC}(t)=[\chi(t)^\top \ \xi_2(t)^\top]^\top$ with dynamics
\begin{equation}
\begin{aligned} \label{dyn3}
    \dot{\chi}&=-\frac{a}{b}\bar{L}(t)\chi+\frac{a}{b}QL(t)\xi_2+Q\bar{\alpha}(e_{PF}) \\
    \dot{\xi}_2&=-\frac{a}{b}L(t)Q^\top\chi-\left(b\mathbb{I}_n-\frac{a}{b}L(t)\right)\xi_2+\bar{\alpha}(e_{PF})-\ddot{\gamma}_d1_n,
\end{aligned}
\end{equation}
where $\bar{L}(t)\trieq QL(t)Q^\top \in \mathbb{R}^{(n-1)\times(n-1)}$. \\% and the spectrum of $\bar{L}(t)$ is the same as that of $L(t)$ without the eigenvalue 0 whose corresponding eigenvector is $1_n$. \\
As a step toward designing a Lyapunov function candidate for \eqref{dyn3}, it is shown that the following auxiliary system
\begin{align} \label{aux}
    \dot{\phi}(t)=-\frac{a}{b}\bar{L}(t)\phi(t), \ \ \ \phi(0)=\phi_0\in\mathbb{R}^{n-1}
\end{align}
is globally uniformly exponentially stable (GUES). Since $Q\in\mathbb{R}^{(n-1)\times n}$ has full rank, there exists $x_0\in\mathbb{R}^n$ such that $\phi_0=Qx_0$. Let $x(t)$ be the solution of \eqref{lem1:equ1}. Then $Qx(t)$ is a unique solution of \eqref{aux}:
\begin{align*}
    \dot{\phi}(t)+\frac{a}{b}\bar{L}(t)\phi(t)&=Q\dot{x}(t)+\frac{a}{b}\bar{L}(t)Qx(t)\\
    &=Q\left(\dot{x}(t)+\frac{a}{b}L(t)Q^\top Qx(t)\right) \\
    &=Q\left(\dot{x}(t)+\frac{a}{b}L(t)x(t)\right)\equiv0,
\end{align*}
where the third equality follows from $L(t)Q^\top Q=L(t)$. Now, one has
\begin{align*}
    \|\phi(t)\|&=\|Qx(t)\|\leq\sqrt{n} \ diam(x(t))\leq \sqrt{n} \ diam(x_0) ke^{-\lambda t} \\
    &\leq \sqrt{n} \bigl(\sqrt{2}\|Qx_0\|\bigr)ke^{-\lambda t}= k_\phi\|\phi_0\|e^{-\lambda t},
\end{align*}
where $k_\phi\trieq\sqrt{2n}k$, the second inequality follows from \eqref{lem1:equ2}, and the remaining inequalities follow from \eqref{lem2:equ}. This GUES of \eqref{aux} provides the basis for constructing a Lyapunov function candidate of \eqref{dyn3} as can be seen next. \\
%Since $\bar{L}(t)$ is continuous for almost all $t\geq0$, uniformly bounded with $\|\bar{L}(t)\|\leq n$, and the system \eqref{aux} is GUES, a similar argument as the one in Theorem $4.12$ in \cite{kha2002} implies that for any constants $c_3$ and $c_4$ satisfying $0<c_3\leq c_4$, there exists a continuously differentiable, symmetric, positive definite matrix $\Psi(t)$ such that
According to Theorem $4.12$ in \cite{Khalil2002}, the GUES of \eqref{aux} suggests that there exists a continuously differentiable, symmetric, positive definite matrix $\Psi(t)$ such that
\begin{align}
    c_1\mathbb{I}_{n-1}\trieq \frac{bc_3}{2an}\mathbb{I}_{n-1}\leq \Psi(t) \leq \frac{k^2_\phi c_4}{2\lambda}\mathbb{I}_{n-1}\trieq c_2\mathbb{I}_{n-1}, \label{p1} \\
    \dot{\Psi}(t)-\frac{a}{b}\bar{L}^\top(t) \Psi(t)-\frac{a}{b}\Psi(t)\bar{L}(t)\leq -c_3\mathbb{I}_{n-1}, \label{p2}
\end{align}
where $c_3$ and $c_4$ are any constants satisfying $0<c_3\leq c_4$.  \\
Now, consider a Lyapunov function candidate for \eqref{dyn3} using $\Psi(t)$ introduced above:
\begin{align} \label{lya}
    V_{TC}(t)=\chi^\top \Psi(t)\chi+\frac{\beta}{2}\|\xi_2\|^2=\bar{\xi}^\top_{TC} W(t)\bar{\xi}_{TC},
\end{align}
where $\beta>0$ and $W(t)\trieq\begin{bmatrix}
\Psi(t) & 0 \\ 
0 & \frac{\beta}{2}\mathbb{I}_n
\end{bmatrix}$. Notice that $V_{TC}$ is bounded by $z^\top M_1z\leq V_{TC}(t)\leq z^\top M_2z$, where $z\trieq[\|\chi\| \ \|\xi_2\|]^\top$ with 
%\begin{align*}
%    z^\top M_1z\leq W(t)\leq z^\top M_2z,
%\end{align*}
\begin{align*} 
M_1\trieq\begin{bmatrix}
c_1 & 0 \\ 
0 & \beta/2
\end{bmatrix} \text{ and }
M_2\trieq\begin{bmatrix}
c_2 & 0 \\ 
0 & \beta/2
\end{bmatrix}.
\end{align*}
The time derivative of \eqref{lya} along the trajectory of \eqref{dyn3} is
\begin{align*}
    \dot{V}_{TC}&=\chi^\top\left(\dot{\Psi}(t)-\frac{a}{b}\bar{L}^\top(t) \Psi(t)-\frac{a}{b}\Psi(t)\bar{L}(t)\right)\chi \\
    &\mathrel{\phantom{=}}-\beta\xi^\top_2\left(b\mathbb{I}_n-\frac{a}{b}L(t)\right)\xi_2 \\
    &\mathrel{\phantom{=}}+\chi^\top\left(2\frac{a}{b}\Psi(t)QL(t)-\beta\frac{a}{b}QL^\top(t)\right)\xi_2 \\
    &\mathrel{\phantom{=}}+\left(2\chi^\top \Psi(t)Q+\beta\xi^\top_2\right)\bar{\alpha}(e_{PF})-\beta\xi^\top_2\ddot{\gamma}_d1_n,
\end{align*}
which leads to 
\begin{align*}
    \dot{V}_{TC}\leq& -c_3\|\chi\|^2-\beta\left(b-\frac{a}{b}n\right)\|\xi_2\|^2 \\
    &+\left(2\frac{a}{b}n\|\Psi(t)\|+\beta\frac{a}{b}n\right)\|\chi\|\|\xi_2\| \\
    &+\left(2\|\Psi(t)\|\|\chi\|+\beta\|\xi_2\|\right)\left(\|\bar{\alpha}(e_{PF})\|+|\ddot{\gamma}_d|\right),
\end{align*}
where we used \eqref{p2}, $\|Q\|=1$, and $\|L(t)\|\leq n$. Applying $\|\Psi(t)\|\leq c_2=\frac{k^2_\phi c_4}{2\lambda}$ in \eqref{p1} yields
\begin{align*}
    \dot{V}_{TC}\leq&-c_3\|\chi\|^2-\beta\left(b-\frac{a}{b}n\right)\|\xi_2\|^2 \\
    &+\left(\frac{a}{b}\frac{nk_\phi^2}{\lambda}c_4+\beta\frac{a}{b}n\right)\|\chi\|\|\xi_2\| \\
    &+\left(\frac{k^2_\phi c_4}{\lambda}+\beta\right)\|\bar{\xi}_{TC}\|\left(\frac{v_{max}}{v_{min}+\epsilon}\|e_{PF}\|+|\ddot{\gamma}_d|\right),
\end{align*}
where $v_{max}=\max_{i}\{v_{i,max}\}$ and $v_{min}=\max_{i}\{v_{i,min}\}$ with $v_{i,max}$ and $v_{i,min}$ being the maximum and minimum achievable speed of the $i$th UAV.
Letting $c_3=c_4$ and $\epsilon>v_{max}-v_{min}$, one obtains
\begin{align*}
    \dot{V}_{TC}&\leq -z^\top Uz+\left(\frac{k^2_\phi c_4}{\lambda}+\beta\right)\|\bar{\xi}_{TC}\|\left(\|e_{PF}\|+|\ddot{\gamma}_d|\right),
\end{align*}
where $z=[\|\chi\| \ \|\xi_2\|]^\top$ and
\begin{align*}
U\trieq\begin{bmatrix}
c_3 & -\frac{1}{2}\left(\frac{a}{b}\frac{nk_\phi^2}{\lambda}c_3+\beta\frac{a}{b}n\right) \\ 
-\frac{1}{2}\left(\frac{a}{b}\frac{nk_\phi^2}{\lambda}c_3+\beta\frac{a}{b}n\right) & \beta\left(b-\frac{a}{b}n\right)
\end{bmatrix}.
\end{align*}
Let  $\lambda_{TC}$ satisfy $\lambda_{TC}\leq\frac{2\lambda}{3k^2_\phi}=\frac{\lambda}{6nk^2}$ and consider 
%which describes the time-coordination convergence rate.
\begin{align*}
    &U-3\lambda_{TC}M_2 \\
    &=
    \begin{bmatrix}
    c_3-\lambda_{TC} \frac{3k^2_\phi}{2\lambda}c_3 & -\frac{1}{2}\left(\frac{a}{b}\frac{nk_\phi^2}{\lambda}c_3+\beta\frac{a}{b}n\right) \\ 
    -\frac{1}{2}\left(\frac{a}{b}\frac{nk_\phi^2}{\lambda}c_3+\beta\frac{a}{b}n\right) & \beta\left(b-\frac{a}{b}n-\frac{3}{2}\lambda_{TC}\right)
    \end{bmatrix}. %\geq 0.
\end{align*}
Note that for a fixed value of $\frac{a}{b}$, all the terms in the above matrix are fixed except for $\beta b$ in the $(2,2)$ element. This is because the values of $k_\phi$ and $\lambda$ are determined by the ratio of $\frac{a}{b}$. Thus,  sufficiently large value of $b$ with a fixed $\frac{a}{b}$ ensures that the above expression is positive semi-definite. \\
Therefore, since $-z^\top Uz\leq-3\lambda_{TC}z^\top M_2z\leq-3\lambda_{TC}V_{TC}$, the derivative of $V_{TC}$ is  bounded above by
\begin{align*}
    \dot{V}_{TC}&\leq -3\lambda_{TC}V_{TC}+\left(\frac{k^2_\phi c_3}{\lambda}+\beta\right)\|\bar{\xi}_{TC}\|\left(\|e_{PF}\|+|\ddot{\gamma}_d|\right) \\
    &\leq-2\lambda_{TC}V_{TC}-\lambda_{TC}\min\{c_1,\beta/2\}\|\bar{\xi}_{TC}\|^2 \\
    &\mathrel{\phantom{\leq}}+\left(\frac{k^2_\phi c_3}{\lambda}+\beta\right)\|\bar{\xi}_{TC}\|\left(\|e_{PF}\|+|\ddot{\gamma}_d|\right).
\end{align*}
Applying Lemma $4.6$ in \cite{Khalil2002} and the state transformation
$\bar{\xi}_{TC}=S\xi_{TC}\trieq
\begin{bmatrix}
b\mathbb{I}_{n-1} & Q \\ 
0 & \mathbb{I}_n
\end{bmatrix}\xi_{TC}$,
we can conclude that
\begin{equation} \label{iss}
\begin{aligned}
    \|\xi_{TC}(t)\|\leq\kappa_1\|\xi_{TC}&(0)\|e^{-\lambda_{TC}t} \\&+\kappa_2\sup_{t\geq0}\left(\|e_{PF}(t)\|+|\ddot{\gamma}_d(t)|\right),
\end{aligned}
\end{equation}
where
\begin{align}
    \kappa_1&\trieq\|S^{-1}\|\sqrt{\frac{\max\{c_2,\beta/2\}}{\min\{c1,\beta/2\}}}\|S\|, \label{kappa1} \\
    \kappa_2&\trieq\|S^{-1}\|\sqrt{\frac{\max\{c_2,\beta/2\}}{\min\{c1,\beta/2\}}}\frac{\frac{k^2_\phi c_3}{\lambda}+\beta}{\lambda_{TC}\min\{c_1,\beta/2\}}. \label{kappa2} 
\end{align}
\end{proof}

\begin{remark}
    The rate of convergence $\lambda_{TC}$ in \eqref{rate} of the time-coordination system is dependent on $T>0$ and $\delta\in(0,T]$, which represent the QoS of the underlying network and  the values of the coordination control gains $a$ and $b$.
\end{remark}

%%%%%%%%%%%%%%%%%%%%%%%%%%%%%%%%%%%%%%%%%%%%%%%%%%%%%%%%%%%%%%%%%%%%%%%%%%%%%%%%
\section{SIMULATION RESULTS} \label{V}
\begin{figure} [h!]
    \centering
    \includegraphics[width = 1.00\linewidth]{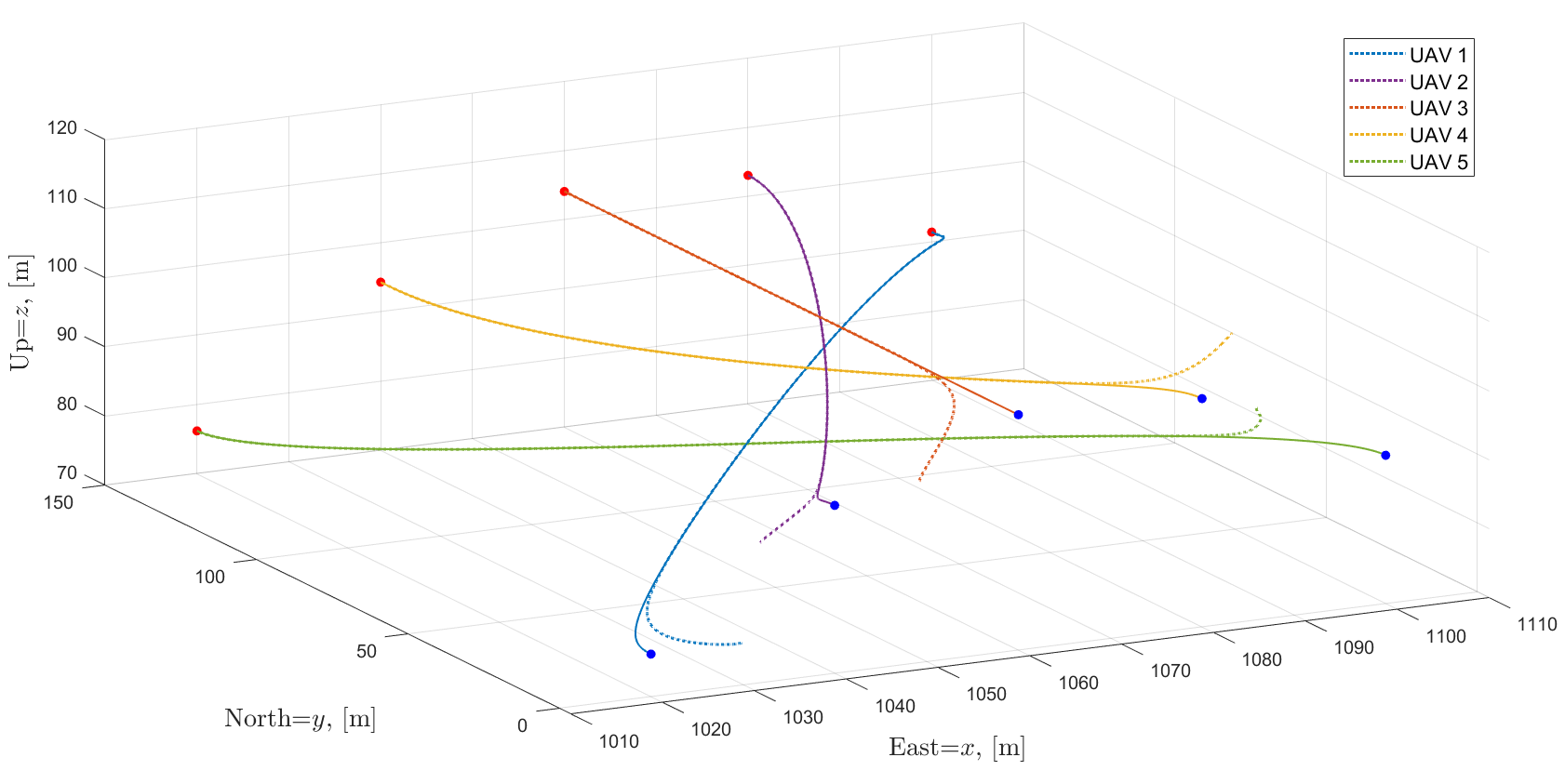}
    \caption{Coordinated path-following of five quadrotor UAVs. Blue dots, the starting points of the desired trajectories, are on $y=0\,m$. Red dots, the end points, are on $y=150\,m$.}
    \label{fig:traj}
\end{figure}

In this section, we consider a coordinated path-following mission to validate the efficacy of the proposed coordination algorithm. The trajectory generation algorithm~\cite{Cichella2021} designs a set of Bezier curves with the mission duration $t_f=19.86\,s$ represented by solid curves in Figure~\ref{fig:traj}. The following specification is considered. The trajectories depart from the plane $y=0\,m$ and simultaneously arrive on the plane $y=150\,m$ exchanging their $(x,z)$ coordinates. The required safe inter-vehicle distance is $10\,m$. 

The quadrotors are initially near the plane $y=0\,m$ with initial path-following errors. The path-following controller~\cite{Cichella2013} allows each quadrotor to track the desired trajectory. This can be confirmed in Figure~\ref{fig:traj}, where the dotted curves depict the paths travelled by the quadrotors. 

\begin{figure} [h!]
     \centering
     \begin{subfigure}[h]{0.1\textwidth}
         \centering
         \includegraphics[width=\textwidth]{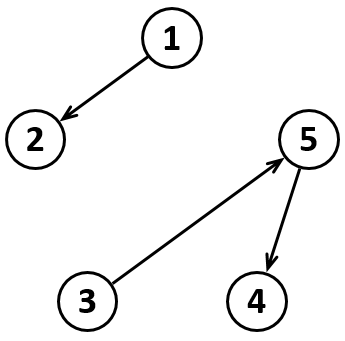}
         \caption{$\mathcal{D}_1$}
         \label{fig:d1}
     \end{subfigure}
     %\hfill
     \hspace{2.5em}
     \begin{subfigure}[h]{0.1\textwidth}
         \centering
         \includegraphics[width=\textwidth]{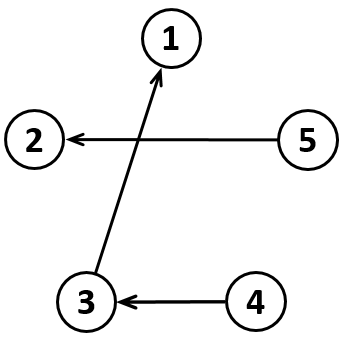}
         \caption{$\mathcal{D}_2$}
         \label{fig:d2}
     \end{subfigure}
     %\hfill
     \hspace{2.5em}
     %\bigskip
     \begin{subfigure}[h]{0.1\textwidth}
         \centering
         \includegraphics[width=\textwidth]{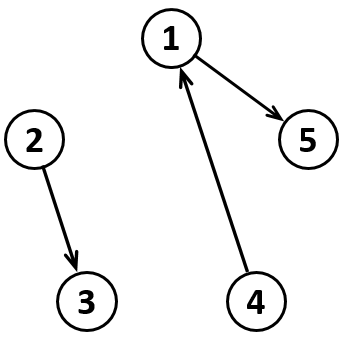}
         \caption{$\mathcal{D}_3$}
         \label{fig:d3}
     \end{subfigure}
     %\hfill
        \caption{Inter-vehicle information flow with digraph topologies.}
        \label{fig:d}
\end{figure}

Suppose that %tight communication bandwidth can support up to three transmission edges at a time. Then, it is reasonable for the quadrotors to transmit their coordination state $\gamma_i(t)$ in a way that 
the topology changes in the order of $\mathcal{D}_1$ $\rightarrow$ $\mathcal{D}_2$ $\rightarrow$  $\mathcal{D}_3$ $\rightarrow$  $\mathcal{D}_1$ $\rightarrow$ $\cdots$, as can be seen in Figure~\ref{fig:d}. With the duration of each topology being $0.03\,s$, the integrated graph $\int_{t}^{t+0.09}L(\tau)d\tau$ $(\forall t\geq0)$ contains $0.03$-spanning tree. In other words, Assumption~\ref{assum3} is satisfied even though $\mathcal{D}(t)$ is not connected at all times. The coordination control gains and a parameter $\epsilon$ in \eqref{alpha} are set to $a=3.75$, $b=4.82$, and $\epsilon=12$, respectively. The initial conditions for the coordination state are $\gamma(0)=0_n$ and $\dot{\gamma}(0)=1_n$.

\begin{figure} [h!]
    \centering
    \includegraphics[width = 1.00\linewidth]{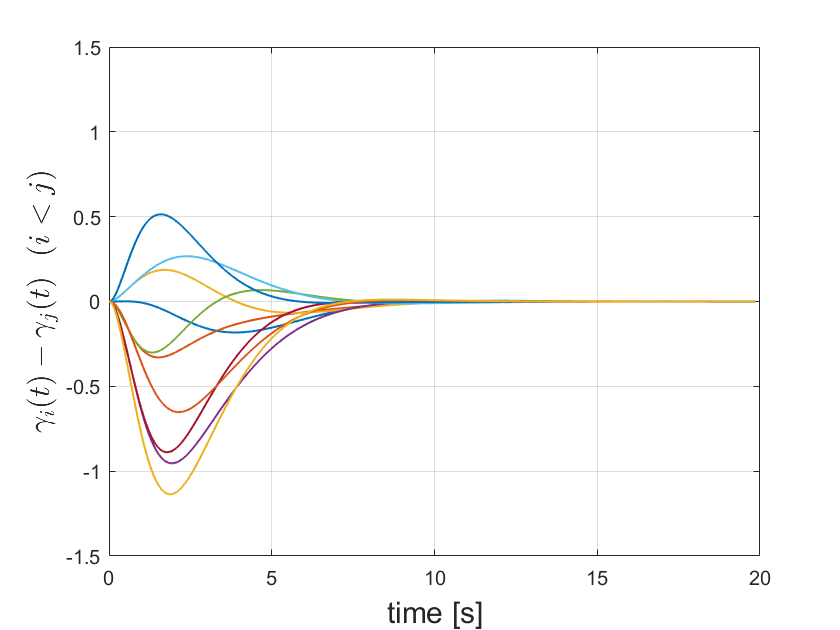}
    \caption{Convergence of the coordination errors $\gamma_i(t)-\gamma_j(t)$ $(i<j)$ to zero.}
    \label{fig:delta_gamma}
\end{figure}

\begin{figure} [h!]
    \centering
    \includegraphics[width = 1.00\linewidth]{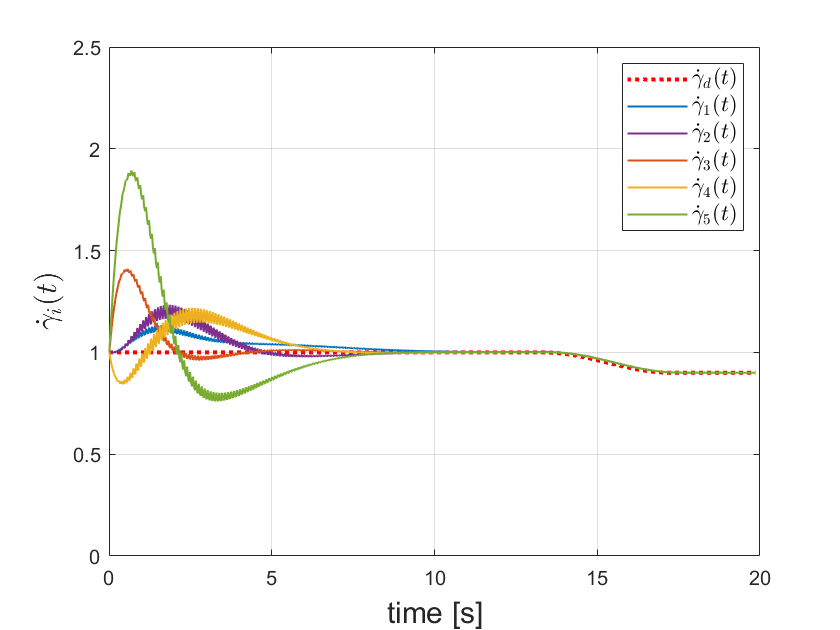}
    \caption{Convergence of the coordination rate $\dot{\gamma}_i(t)$ to the desired mission rate $\dot{\gamma}_d(t)$.}
    \label{fig:gamma_dot}
\end{figure}

\begin{figure} [h!]
    \centering
    \includegraphics[width = 1.00\linewidth]{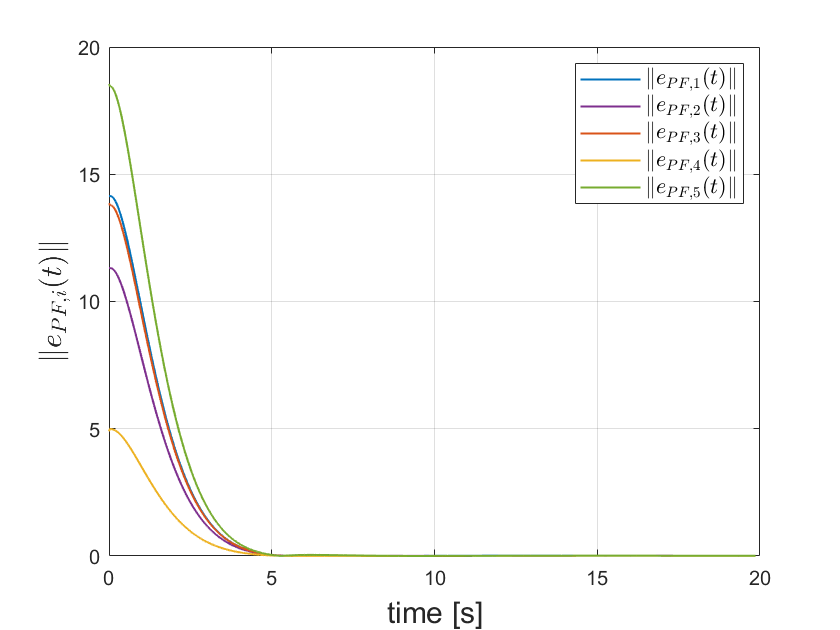}
    \caption{Path-following errors}
    \label{fig:e_pf}
\end{figure}

 Figures~\ref{fig:delta_gamma}, \ref{fig:gamma_dot}, and \ref{fig:e_pf} illustrate how the coordination controller \eqref{dyn1} operates. At time $t=0\,s$, the quadrotors $1$, $2$, $3$, and $5$ lie ahead of the trajectories and the quadrotor $4$ is in the other case, which causes $\bar{\alpha}_i(e_{PF,i}(t))$ to be positive for $i=1,2,3,5$ and negative for $i=4$, thereby accelerating and decelerating $\gamma_i(t)$, respectively, as seen in Figure~\ref{fig:gamma_dot}. Thus, it confirms that $\bar{\alpha}_i(e_{PF,i}(t))$ reduces the size of $\|e_{PF,i}(t)\|$ helping each quadrotor stay inside the region of attraction. However, the increases of the coordination errors for the first few seconds in Figure~\ref{fig:delta_gamma} indicate that it comes at the cost of increased discoordination. In Figure~\ref{fig:e_pf}, as $\|e_{PF,i}(t)\|$ decreases by the collaborative efforts of $\bar{\alpha}_i(e_{PF,i}(t))$ and the path-following controller, the impact of $\bar{\alpha}_i(e_{PF,i}(t))$ on the coordination controller \eqref{dyn1} decreases and the first two terms in the right hand side of \eqref{dyn1} achieve the coordination objectives \eqref{obj1} and \eqref{obj2}. Figure~\ref{fig:delta_gamma} clearly shows that $\gamma_i(t)-\gamma_i(t)$ converges to zero achieving the first coordination objective \eqref{obj1}. In Figure~\ref{fig:gamma_dot}, we can see that $\dot{\gamma}_i(t)$ tracks $\dot{\gamma}_d(t)$ that varies from $1$ to $0.9$ at about $t=15.3\,s$, which illustrates that the second coordination objective \eqref{obj2} was met. The nonsmooth evolution of $\dot{\gamma}_i(t)$ for the first few seconds in Figure~\ref{fig:gamma_dot} is due to the following fact. 
 The switching nature of the network topology in Figure~\ref{fig:d} leads to switching of the Laplacian $L(t)$ in \eqref{collective_dyn}. It yields the nonsmooth evolution of $\dot{\gamma}_i(t)$ for the first few seconds.
 %We note that the acceleration of $\gamma_i(t)$'s for the first few seconds had the mission unfold faster than planned. To fix it, the desired mission rate $\dot{\gamma}_d(t)$ was adjusted from $1$ to $0.9$ around $t=15.3\,s$. With this adjustment, the fleet simultaneously arrive at their destinations as planned at $t=19.86\,s$. 
 In the case when the quadrotors are deviated from the desired trajectories by disturbances such as head winds or tail winds, they recover the coordination automatically in the same manner.

%\begin{figure}
%     \begin{subfigure}[b]{0.24\textwidth}
%         \centering
%         \includegraphics[width=\textwidth]{gamma.png}
%         \caption{}
%         \label{fig:} 
%     \end{subfigure}
%     \hfill
%     \begin{subfigure}[b]{0.24\textwidth}
%         \centering
%         \includegraphics[width=\textwidth]{delta_gamma.png}
%         \caption{}
%         \label{fig:}
%     \end{subfigure}
%     \vfill
%     \begin{subfigure}[b]{0.24\textwidth}
%         \centering
%         \includegraphics[width=\textwidth]{gamma_dot.png}
%         \caption{}
%         \label{fig:}
%     \end{subfigure}
%     \hfill
%     \begin{subfigure}[b]{0.24\textwidth}
%         \centering
%         \includegraphics[width=\textwidth]{e_pf.png}
%         \caption{}
%         \label{fig:}
%     \end{subfigure}
%        \caption{}
%        \label{fig:}
%\end{figure}
%%%%%%%%%%%%%%%%%%%%%%%%%%%%%%%%%%%%%%%%%%%%%%%%%%%%%%%%%%%%%%%%%%%%%%%%%%%%%%%%
\section{CONCLUSION} \label{VI}
This paper presents a distributed time-coordination algorithm that solves the Time Coordinated Path Following Problem under a very general   connectivity assumption on the underlying communication network, namely that it is represented by a directed graph that is connected in an integral sense. Under this assumption, we showed using Lyapunov analysis that the distributed time-coordination algorithm guarantees exponential convergence of the coordination error to a neighborhood of zero and thus solves the TCPF problem. Simulation results validated the efficacy of the proposed algorithm. Future work will analyze the stability of the proposed algorithm with time-delayed communication. %in the presence of quantized information flow and time-delayed communication.

%%%%%%%%%%%%%%%%%%%%%%%%%%%%%%%%%%%%%%%%%%%%%%%%%%%%%%%%%%%%%%%%%%%%%%%%%%%%%%%%
%\section{ACKNOWLEDGMENTS}

%%%%%%%%%%%%%%%%%%%%%%%%%%%%%%%%%%%%%%%%%%%%%%%%%%%%%%%%%%%%%%%%%%%%%%%%%%%%%%%%

\bibliographystyle{ieeetr}
\bibliography{references}

\end{document}

%% file: definition.tex
\usepackage{xcolor}
\usepackage{float}
\usepackage{amsmath}

\usepackage{amsthm,amssymb,graphicx,url}
\let\labelindent\relax 
\usepackage{enumitem}
\usepackage{booktabs}
\usepackage{caption}
\usepackage{subcaption}

\usepackage{appendix}

\usepackage{wrapfig}
\usepackage{hyperref}%[bookmarks=true]
\usepackage{soul}  % for using strike through
\usepackage{cleveref}
\usepackage{bm}
\usepackage{multicol}
\usepackage{mathtools}
\usepackage{cite}
\usepackage{graphicx}

\crefname{equation}{}{} %skip "eq" or "eqs". 
\crefname{assumption}{Assumption}{}
\crefname{table}{Table}{} 
\crefname{figure}{Fig.}{}
\crefname{section}{Section}{}
\crefname{remark}{Remark}{}
\usepackage{algorithm,algorithmicx} %linesnumbered is an option [ruled], algorithm is a warpper
\usepackage{algpseudocode} % layout for algorithmicx
\newlength\myindent
\def\trieq{\triangleq}

\newtheorem{theorem}{Theorem}
\newtheorem{lemma}{Lemma}
\theoremstyle{definition}  
\theoremstyle{definition} \newtheorem{assumption}{Assumption}
\theoremstyle{remark}  \newtheorem{remark}{Remark}